# How reliable are remote sensing maps calibrated over large areas? A matter of scale?

Luigui Andrey Ramirez Parra[1,2], Jean-Pierre Renaud[2,3], Cédric Vega[3]•

[1] Université de Lorraine, INRAE, UR 1138 BEF, F-54280 Champenoux, France. Luigui-Andrey.Ramirez.Parra@inrae.fr

[2] Office National des Forêts, Département RDI, 5, rue Girardet - CS 65219, 54052 Nancy Cedex, France. jean-pierre.renaud-02@onf.fr;

[3] Laboratoire d'Inventaire Forestier, ENSG, IGN, INRAE, 14 rue Giradet, Nancy, 54000, France. Cedric.vega@ign.fr

*Key message*

Models established at a broad sampling resolution should be considered with care when applied locally as they may present substantial biases.

**Abstract**

*Context*:

Remote sensing data are increasingly available and frequently used to produce forest attributes maps. The sampling strategy of the calibration plots may directly affect predictions and map qualities.

*Aims*:

Evaluate models transferability at different spatial scales according to the sampling efforts and the calibration domain of these models.

*Methods:*

Forest inventory plots from locals and regionals networks were used to calibrate randomForest (RF) models for stand basal area predictions. Auxiliary data from ALS flights and a Sentinel-2 image were used. Model transferability was assessed by comparing models developed over a given area and applied elsewhere. Performances were measured in terms of precision (RMSE and bias), coefficient of determination ($R^2$) and the proportion of extrapolated predictions. Regional networks were also thinned to evaluate the effect of sampling efforts on models' performances.

*Results*:

Local models showed large bias and extrapolation issues when applied elsewhere. Local issues of regional models were also observed, raising transferability and extrapolation concerns. An increase in sampling efforts was shown to reduce extrapolation issues.

*Conclusion:*

The outcoming results of this study underline the importance of considering models' validity domain while producing forest attribute maps, since their transferability is of crucial importance from a forest management perspective.

1. **Introduction**

The use of remote sensing data for forest management is rapidly increasing and widely implemented. Since the early 2000, Airborne Laser Scanning (ALS) and 3D remote sensing data have been used in forestry to assist forest attributes estimations (Gregoire et al. 2011; Kangas and Maltamo 2006; Vega et al. 2021). It has also been used to identify wind damages (Renaud et al. 2017), stand structure, site index (Véga 2006) or even habitats. It proved to be useful to produce Digital Terrain Models (DTM) and help archaeological investigations (Georges-Leroy et al. 2011). Nowadays, countries such as Finland, Sweden, and Norway among others, are producing forest attributes maps at a national or departmental level based on ALS data.

In France, several ALS flights have been performed so far, but generally covering limited areas such as forest stands or small territories. The resulting maps are then produced from models developed for each individual acquisition. However, recently, the French government have initiated a program intending to cover the whole metropolitan areas and overseas territories with a high density ALS (LiDAR HD) acquisition aimed to end by 2025 (National Institute of Geographic and Forest Information, IGN, 2020). Such a program open perspectives for upscaling models and estimation appraoches from local to national level.

Since the early 1990's, multisource forest inventory approaches have been developed, combining remote sensing data and field inventory plots, that improve forest attributes assessments (Holmgren and Thuresson 1998; Tokola and Heikkilä 1997; Tomppo and Katila 1991; White et al. 2016). Among its benefits, use of remote sensing data as an auxiliary information source is a cost-effective means to improve precision of forest inventories, which leads to better planning or continuous resource assessment (Poso et al. 1999; Tokola et al. 1996). Moreover, the provision of maps for forest management based on multisource inventories can assist decision-making processes, reduce future inventories requirements, and inclusively add information to unsampled areas (Fekety et al. 2015). However, when models are made using NFI plots, their calibration domain tends to be sparse and their predictions for small areas dedicated to forest management might be less precise (Hsu et al. 2020; Vega et al. 2021, De Lera Garrido et al. 2023).

The resulting maps products, which represent a simplified version of the reality, translate the common features of forests within an area of interest (AOI). However, it is important to know that the outcoming maps may have a limited calibration domain and at the local level, may be significantly biased (Sagar et al. 2022; Kangas et al. 2023; Ståhl et al. 2024). The way the NFI sampling scheme is designed, the spatial distribution and density of the plots, directly affect models and map qualities, particularly for forest management (Poso et al. 1999). Model predictions include uncertainties, often summarised as models' root mean square error (RMSE) (James et al. 2013). When a model is applied to an external dataset, in other locations, the auxiliary data space of the pixel population may be different from the one used to calibrate it and predictions may be exposed to extrapolation issues leading potentially to unreliable management decisions (Bartley et al. 2019; Hsu et al. 2020; Sagar et al. 2022).

Model calibration is frequently done by using datasets from National Forest Inventories that represent large domain areas (Tomppo et al. 2008; Tomppo and Katila 1991; De Lera Garrido et al. 2023). Although these models perform globally well, they may produce locally unreliable predictions, especially if the auxiliary data are locally out of the model calibration domain. On the other hand, it is tempting as well to consider local inventory networks (usually done for forest management) to calibrate statistical models and assess to what extent they can remain reliable over larger areas from spatially continuous predictors (Meyer and Pebesma 2021). This rises the idea of "model transferability" with a sufficient sampling effort, and precision required to produce reliable maps for management.

So far, several studies involved models developed for local application exclusively. However, with the increasing availability of data over large areas, the aspect of model transferability become a major concern (Yates et al. 2018). According to van Ewijk et al. (2020), the possibility of transferring models to new locations may bring improvements in monitoring and resource assessment, but requires special attention to the calibration and performance of the models implemented. There are several features according to Yates et al. (2018) that account for the transferability of models, such as the resolution of the data, the spatio-temporal bias, and the complexity of the model. The mentioned factors are determinant in the model adjustment, on how it predicts through unknown space when transferred, how much variability of the target attribute can be explained. The previous idea is linked to the importance of the model calibration domain when maps derivatives are being used. Brooks et al. (1988) define the space where the model's predictions are valid as '*validity domain*'. When predictions are done outside this domain with insufficient samples, the transferability might be compromised and lead to extrapolation issues. Furthermore, as the number of predictors involved in the predictions increase, so does the complexity of the validity domain and its dimensionality. This situation often led to overfitting problems and to the curse of dimensionality, which could seriously impair models' reliability (Sagar et al. 2022).

One way to characterize the validity domain and diagnose extrapolation in map products is to use a convex hull approach (Brooks et al. 1988; Renaud et al. 2022). By doing so, a new prediction is tested to determine whether it belongs to a model's calibration range in the auxiliary data space. With the ever increasing availability of remote sensing data, models are often build over large areas (Guerra-Hernández et al. 2022). Regardless of the fact that these models could be globally unbiased, map accuracy at local level is a major concern especially for forest management (Hsu et al. 2020; Kangas et al. 2023; Ståhl et al. 2024).

In this regard, the main objectives of this study were: to evaluate how models developed from plots sampled over a large territory are producing local biases; and conversely how models developed over a given forest behave when applied to larger territories or adjacent forests. A "Regional model" refers to a model developed from extensive data across the whole AOI such as NFI or large scale Management Forest Invenotry (LSMFI) (e;g. a forest Observatory network). A "local model" stands for a model developed from a local network of field management plots (MFI) on specific forest stands.

Specific objectives were :

- To evaluate the transferability of models : impact on precision and accuracy indicators ($R^2$, RMSE, Bias) in both upscaling and downscaling applications of models.
- To evaluate for regional networks, how these indicators, as well as the proportion of extrapolated predictions are influences by the sampling intensity, via a subsampling approach.
- To evaluate the level of extrapolation observed when regional models are applied locally, and report these events on map products for forest management.

2. **Material and methods**

*2.1. Study area*

The study area is located in North-Eastern France and comprises the Northern part of the Vosges mountains (Figure 1). The area is characterized by constrasted landscapes, including lowlands, hills, and mountains, spreading from 400 m to 1,424 m above see level. The mountainous part is characterized by a continental climate, with mean annual temperature ranging from 6 to 10 °C according to altitude and orientation, and annual precipitations varying from 900 to 2,000 mm with mean annual value of 1116 mm. It is coverted at 75 % by forests dominated by harwoods species, up to 500 m, and a mixture of beech (*Fagus Sylvatica* L.), fir (*Abies alba* Mill.) and spruce (*Picea abies* (L.) H.Karst) in higher elevations. The lowland and hilly areas formed the peri-vosgian region. The forest occupied 44 % of the area and are most diversified in terms of composition and structure, with mixture of beech and oak (*Quercus petraea* (Mattus.) Liebl.), and birch (*Betula* spp.), pine (*Pinus sylvestris* L.), chestnut (*Castanea sativa* Mill.), among others (https://inventaire-forestier.ign.fr/).

## 2.2. Field data

Field data included both NFI and MFI data, representing a total of 2,828 plots. NFI data consisted in 585 plots surveyed from 2016 to 2020. The NFI sampling design relies on a systemetic sampling desing based on a 1 km grid divided into two 5 years samples, from which annual samples are randomly selected. Field measurements are done for an annual sub-sample of ~ 6500 plots which consist in concentric circular plots of 6, 9, 15 and 25 m radii (IGN, 2022). The later plot is mainly used to describe stand structure. Tree measurements are done in the three smaller circles, according to tree diameter classes, [7.5 – 22.5 cm , 22.5 – 47.5 cm and 47.5 – 67.5 cm respectively]. MFI data were acquired by the National Forest Office (ONF) and consisted into 2 datasets. The first dataset was acquired in the framework of the establisement of forest observatories, following a systematic sample design and consisted in 198 plots measured on 2021. The second dataset consisted in 2045 plots surveyed in 5 different forests (i.e. Deodatie, Deux lacs, Donon, Gueberschwihr, Mouterhouse) using specific sampling designs. For the inventory plots in the forests of Gueberschwhir and Deux Lacs, tree measurements were done by the angle count method (Duplat and Perrotte 1981). For the other MFI, field measurements were done in 15 m radius plots for trees having a diameter at 1.3m of 17.5 cm and larger (DBH >= 17.5 cm).

In the framework of this study, the targeted forest attribute was the plot basal area (Ba, m²/ha). To allow comparisons between NFI and MFI data, plot level computations were harmonized. Only the NFI plots fully included in a stand were considered, to match with MFI sampling design. Plot level computation were done according to the NFI estimation process, for the subset of trees trees having a DBH >= 17.5 cm.

Table 1: Summary of plot characteristics, where MFI = Management Forest Inventory, LSMFI = Large Scale Management Forest Inventory,  Ba = Basal area, Sd = Standard deviation, N° plots = number of plots per forest types.

| Forest | Type of Inventory | Sampling design | Forest Area(ha) | Year of the field survey | N° of plots | Sampling intensity (ha/plot) | Ba(m²/ha) | | N° Plots | | |
|---|---|---|---|---|---|---|---|---|---|---|---|
| | | | | | | | Mean (m²/ha) | Sd | Broadleaves | Mixed | Conifers |
| **Deodatie** | MFI | Random | 43750 | 2019 | 192 | 228 | 27.2 | 13.9 | 31 | 40 | 121 |
| **Deux lacs** | MFI | Systematic | 638 | 2020 | 346 | 2 | 27.0 | 12.7 | 0 | 17 | 329 |
| **Donon** | MFI | Systematic | 11984 | 2020-2021 | 561 | 21 | 26.8 | 9.8 | 119 | 77 | 365 |
| **Gueberschwihr** | MFI | Systematic | 416 | 2020-2021 | 402 | 1 | 28.2 | 8.6 | 110 | 126 | 166 |
| **Mouterhouse** | MFI | Systematic | 5356 | 2019 | 544 | 10 | 23.6 | 8.3 | 280 | 89 | 175 |
| **Observatory** | LSMFI | Systematic | 254277 | 2021 | 198 | 1284 | 25.6 | 11.9 | 61 | 52 | 85 |
| **NFI** | NFI | Systematic | 254277 | 2016-2020 | 585 | 435 | 30.6 | 13.8 | 181 | 145 | 259 |

## 2.3. *Auxiliary data and processing*

Auxiliary data from ALS, Sentinel 2 and a forest mask (BD Forêt v2) were used. The ALS data was acquired from IGN, and supplied in the form of 1 squared km las files of classified point clouds, and the outer limit of the tiles was used to define AOI. The ALS data consisted in three flight campaings over the AOI (Figure 1**Erreur ! Source du renvoi introuvable.**). The South-Eastern (SE) area survey consisted of 626 tiles during winter 2021 with a Leica ALS70 at 1,797 m above ground level, pulse and scanning frequencies of 309 Hz and 35 Hz, a maximum scan angle of 20°. The Northern (N) area survey consisted of 2424 tiles aquired in 2020 and the South-Western (SW) area survey consist of 3428 tiles also acquired in 2020. The output mean point density over the whole AOI was 5 pts/m².

Point cloud processing was done by IGN using Terrascan software (www.terrasolid.com). Ground classified points were used to generate a 1 m resolution digital terrain model (DTM) through inverse distance weighted interpolation. The elevation of the points above the ground was converted into height by subtracting the corresponding values from the DTM. Subsequently, the normalized point cloud data was used to generate a 1 m canopy height model (CHM) by selecting the maximum point height per pixel. In cases where pixels did not have values, interpolation was performed by averaging the values of the eight neighbouring pixels. Finally, the extraction of ALS metrics for the AOI was done according to Vega et al. (2016).

A Sentinel 2 image was obtained from THEIA platform (https://www.theia-land.fr/) with processing level 2B, which included topographic and atmospheric corrections (Hagolle et al. 2010) . The image was acquired in 2019 under cloud free conditions. Spectral bands having a 20 m spatial resolution were resampled to 10 m resolution using R library raster. Only pixels included in the AOI were selected for further processing. The BD Forêt 2.0 database of IGN was used to determine forest types. Forest type was considered as a factor within the models, corresponding to the type of vegetation at the plot (or pixel) level (i.e., Broadleaves, Mixed, or Conifers).

## 2.4. *Study Area*

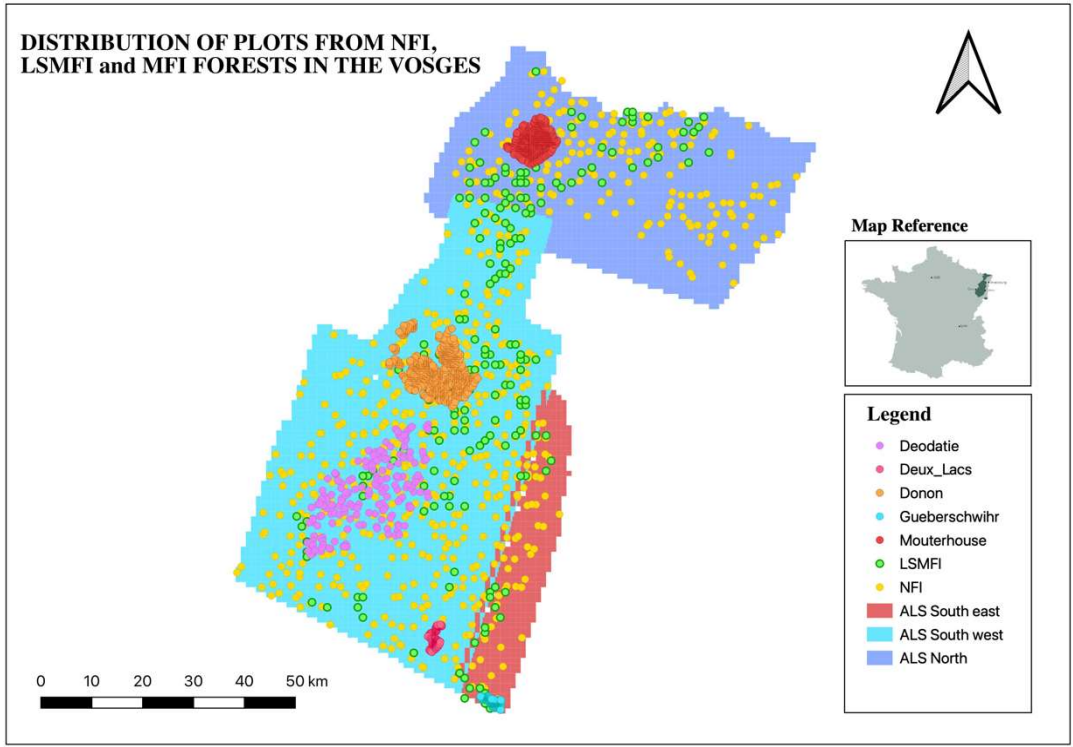

*Figure 1: Study area and localisation of plots. Where each ALS flights are as follows: In orange the ALS flight on the South east, In blue the ALS flight on the South west, and in purple the ALS flight from the North; with each set of field plots within the AOI from MFI and NFI in different colours (see legend).*

## 2.5. *Modelling approach*

In this study, a modeling approach using Random Forest (RF, Liaw and Wiener 2002) was used to predict Ba ($m^2$/ha). Figure 2 illustrates the process behind the division and selection of the samples. For each dataset from MFI-LSMFI and NFI, 80% of the plots was selected for training, and the remaining 20% for testing. Then, the training dataset was subdivided in calibration (80% of plots) and validation sets (20% of the plots) to select and apply the preliminary models from the ALS metrics, the Sentinel bands and forest types. In this study, all the returns above 2 meters were selected from the ALS as auxiliary data. As seen on Figure 2 the validation/testing sets are used to evaluate the model results obtained from the training (calibration) sets.

## 2.6. *Variable selection*

The rationale behind variable selection is to obtain a good model in further steps, reduce overfitting and remove variables that add unnecessary complexity to the model (James et al. 2013). Complex models could lead to overfitting, producing biased predictions, noise or limiting transferability (Babyak 2004; Yates et al. 2018). The most important predictors were identified for each dataset. The auxiliary data consists in a set of predictors that were extracted from the ALS flight's metrics corresponding to the Canopy height model (CHM) described by (Véga et al. 2016), as well as other metrics calculated by the lidR package (Roussel et al. 2020). Sentinel-2 metrics and Forest types from BD forêt version 2.0 were also extracted at the plot level.

The selection of auxiliary data is based on a Lasso regression. It consists in a shrinkage of the estimate coefficients towards zero; where the least important predictors are forced to be exactly equal to zero, and then removed from the model (James et al. 2013). The cv.glmnet function of the glmnet package (Tay et al. 2023) was used and the non-zero metrics were retained. They were then further compared with their importance values based on the importance function of the randomForest package (Liaw and Wiener 2002). The variables retained in the models are in Table 2.

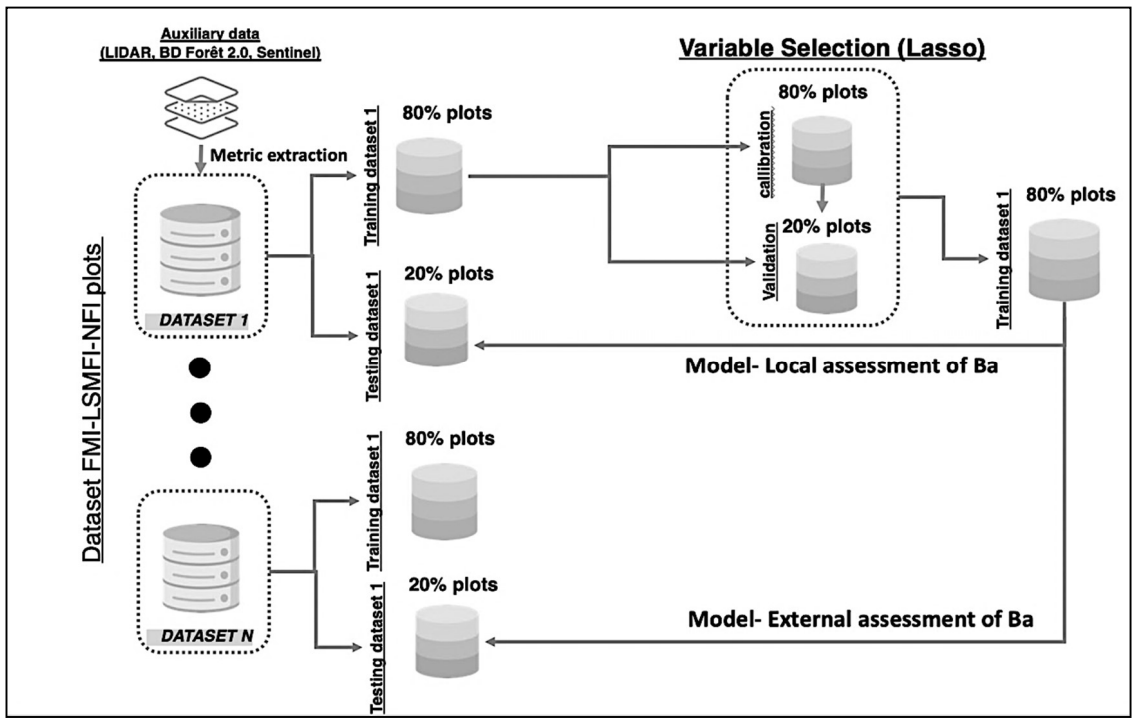

*Figure 2: Treatment of the data for the model definition and assessment with the MFI, LSMFI, and NFI datasets.*

### 2.7. *Evaluation of model transferability and performance*

Once the selection of predictors is done, model transferability is assessed as follows: 1) against its testing dataset (20% of the plots) for a local assessment, and 2) against the testing datasets from the remaining forests (external assessment, Figure 2). This guarantee the independence of each model testing, applying datasets that were never used for model training. The main indicators used to define the performance of the testing were:

the R²:

$$R^2 = 1 - \frac{\sum_{i=1}^{n}(y_i - \hat{y}_i)^2}{\sum(y_i - \bar{y})^2}$$

the Root Mean Square Error (RMSE):

$$RMSE = \sqrt{\frac{\sum_{i=1}^{n}(y_i - \hat{y}_i)^2}{n-1}}$$

the mean bias:

$$Mean\ bias = \frac{\sum_{i=1}^{n}(y_i - \hat{y}_i)}{n}$$

where **n** is the number of tested plots, $y_i$ the basal area (m²/ha) observed from field surveys, $\bar{y}$ its mean, and $\hat{y}_i$ the predicted basal area (m²/ha) from the model.

### 2.8. *Evaluating the impact of sampling effort*

The evaluation of the sampling effort was done for regional networks of NFI and LSMFI plots. It started by building a grid for the extent of the AOI to sample one plot per grid cell. Then, five grid resolutions were implemented ranging between (2, 4, 6, 10, 20 km) (Figure 3). Finally, an iteration was performed for each grid resolution (4, 4, 5, 6, 7 times respectively) due to the low density of plots per cell and also to reduce the computational effort. For each iteration sample, a random plot was selected per cell and used to evaluate the goodness of fit of the Ba indicators (R²; RMSE; Bias).

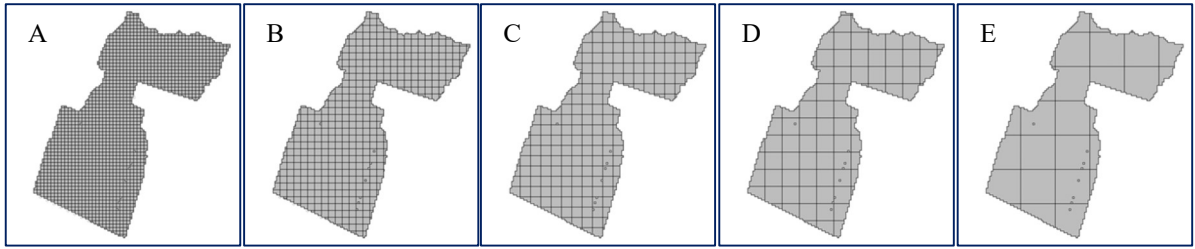

*Figure 3: Example of the grid resolution adjustment to evaluate sampling effort, where: A) Resolution of 2km, B) Resolution of 4km, C) Resolution of 6 km, D) Resolution of 10km, and E) Resolution of 20 km*

### 2.9. *Evaluation of extrapolation risk*

The evaluation of the extrapolation risk was done with a convex hull approach using the R package "geometry" (Habel et al. 2023) and the Inhull function. The "yaImpute" package (Crookston and Finley 2008) was used to determine distances of the extrapolated pixels to the nearest calibration plot, as well as to compute the mean interpolation distance (MCD), consisting in the mean euclidean distance between each pair of calibration plots in the auxiliary space (Renaud et al. 2022). The convex hull was built using the calibration auxiliary data for each model. It gave the calibration envelop, allowing to characterize pixel predictions of the AOI as being inside or outside the calibration domain (extrapolated). Extrapolated pixels were further classified as being "Near" or "Far" from the calibration domain based on their distance to the nearest calibration plot. When this distance was less or equal to MCD, it was classified as "Near" to the calibration domain, and when the distance was larger than MCD it was considered "Far" and at a substantial risk of unreliable prediction (Sagar et al. 2022).

### 2.10. *Map production*

To create maps, RF models (Table 2) were used to get pixels prediction over the AOI, using the auxiliary data. Similarly, the extrapolation level maps ("Near" and "Far") were built to identify extrapolation risk according to the classification previously described (

Figure 4). Raster maps of Ba (m$^2$/ha) were produced at a pixel resolution of 30m.

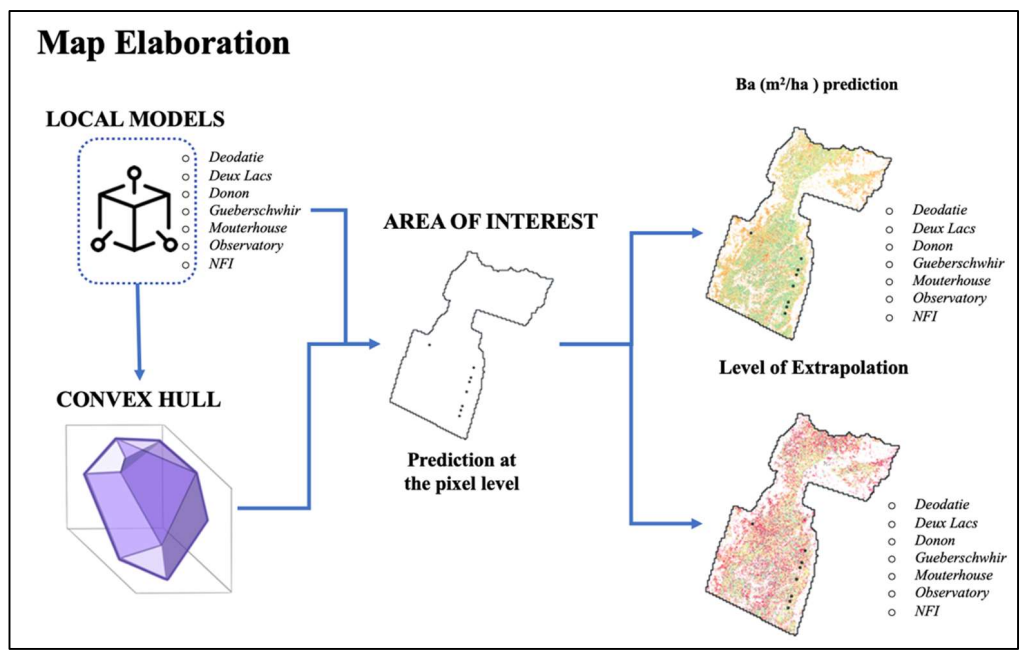

*Figure 4: Sequence in the map elaboration process. Starting from the local models until the map products. Convex hull is schematically represented here as a 3D space of the auxiliary data.*

## 3. Results

### 3.1. *Selected variables and Implemented Models:*

Table 2 lists the retained models for each specific dataset and their auxiliary variables, for predicting Ba. The most frequently selected and important predictors for the Ba (m$^2$/ha), across all different datasets correspond to the ALS metric "VolIn" which refers to the inner canopy volume of the plot, the gap ratio which accounts for the gap surface within the plot (Vega et al. 2016), and the Band 11 and Band 12 SWIR (Short Wave Infrared) which are frequently considered for Ba predictions (Haley *et al*, 2005).

*Table 2: Model retained for each data sources (Forest).*

| Forest | Retained Model |
|---|---|
| Deodatie | Ba ~ VolIn + B11 + Canopy closure + Z mean + Forest type |
| Deux Lacs | Ba ~ VolIn + B11 + Zq 5 + Zq 10 + Forest type |
| Donon | Ba ~ VolIn + B11 + Gap ratio + p2th +B12 + Forest type |
| Gueberschwhir | Ba ~ VolIn + B11 + Gap ratio + Zq 30 +B12 + Forest type |
| Mouterhouse | Ba ~ VolIn + B11 + Gap ratio + Zq 50 + Forest type |
| LSMFI | Ba ~ VolIn + B11 + Gap ratio + Forest type |
| NFI | Ba ~ VolIn + B11 + Gap rqtio + Z mean + Forest type |

### 3.2. *Model's transferability:*

Table 3 illustrates the performance of the models applied. when used locally, the models built on Deodatie, Donon or Mouterhouse produced the highest percentage of explained variability ($R^2$) for Ba, with 87%, 80% and 81% respectively. They also had low RMSE values (respectively 5.63, 4,26 and 3,75m$^2$/ha). Similarly, the models corresponding to the NFI and LSMFI plots, when used over their own datasets, explained slightly less (i.e., 78% and 67% ) of the Ba variability, and had slightly larger RMSE ( 6,05 and 7,6 m$^2$/ha respectively). When applied over foreign datasets, a strong reduction in models' predictive power is observed ($R^2$ drop below 50%). It is also noteworthy to observe that in some situations, negative $R^2$ were acheived (e.g. for the Doedatie model applied to the Mouterhouse forest), indicating that in these situations the use of the model was worst that using a simple mean of field observations.

*Table 3 : Model performances tested over the different AOI considered. $R^2$, RMSE (m2/ha) and Bias of the models at local/global levels.*

| Model / Forest | Deodatie | | | Deux lacs | | | Donon | | | Gueberschwihr | | | Mouterhouse | | | Onf Observatory | | | NFI | | |
|---|---|---|---|---|---|---|---|---|---|---|---|---|---|---|---|---|---|---|---|---|---|
| Performance | R2 | RMSE | Bias | R2 | RMSE | Bias | R2 | RMSE | Bias | R2 | RMSE | Bias | R2 | RMSE | Bias | R2 | RMSE | Bias | R2 | RMSE | Bias |
| Deodatie | **0,87** | **5,63** | **0,55** | 0,34 | 12,80 | -2,31 | 0,49 | 11,30 | 3,63 | 0,43 | 11,90 | -0,87 | 0,32 | 13,00 | 6,88 | 0,52 | 10,20 | 0,28 | 0,21 | 14,10 | -6,26 |
| Deux lacs | 0,15 | 11,90 | 5,15 | **0,57** | **8,46** | **0,49** | 0,15 | 11,90 | 6,82 | 0,46 | 9,47 | 2,13 | -0,17 | 14,70 | 9,84 | 0,51 | 8,56 | 1,70 | 0,40 | 9,99 | -3,53 |
| Donon | -0,23 | 10,50 | -6,60 | -0,01 | 9,54 | -5,34 | **0,80** | **4,26** | **-0,12** | 0,35 | 7,64 | -2,67 | 0,52 | 6,57 | 1,28 | 0,46 | 7,00 | -2,32 | -0,01 | 9,51 | -7,31 |
| Gueberschwihr | 0,09 | 8,72 | -2,51 | 0,30 | 7,65 | -0,89 | 0,38 | 7,22 | 2,82 | **0,79** | **4,21** | **0,37** | 0,08 | 8,78 | 5,19 | 0,26 | 7,24 | 1,09 | 0,28 | 7,77 | -3,46 |
| Mouterhouse | -0,82 | 11,80 | -8,99 | -0,27 | 9,80 | -6,97 | 0,56 | 5,79 | -1,17 | 0,16 | 7,98 | -4,02 | **0,81** | **3,75** | **0,24** | 0,25 | 7,10 | -2,46 | -0,27 | 9,82 | -7,34 |
| LSMFI | 0,30 | 11,00 | -4,07 | 0,22 | 11,50 | -2,97 | 0,55 | 8,75 | 2,14 | 0,50 | 9,22 | -1,15 | 0,47 | 9,54 | 4,00 | **0,76** | **6,20** | **0,27** | 0,43 | 9,85 | -4,78 |
| NFI | 0,20 | 12,00 | 3,28 | 0,42 | 10,20 | 2,74 | 0,13 | 12,50 | 7,85 | 0,38 | 10,60 | 2,66 | -0,10 | 14,10 | 9,65 | 0,33 | 11,00 | 5,32 | **0,68** | **7,60** | **0,52** |
| mean | -0,05 | 10,99 | -2,29 | 0,17 | 10,25 | -2,62 | 0,37 | 9,58 | 3,68 | 0,38 | 9,47 | -0,65 | 0,18 | 11,12 | 6,14 | 0,39 | 8,52 | 0,60 | 0,17 | 10,17 | -5,45 |

In terms of precision, the less precise model (i.e. with the highest RMSE, tested over its own data) was the one developed over the Deux lacs forest (8,46 m$^2$/ha). Moreover, when models were applied to other forests, RMSE increases drastically in most of the cases. It reached the highest value (14,7m$^2$/ha) when the Mouterhouse model was applied to the Deux lacs dataset. In terms of bias, similarly, local models had neglectable bias when applied over their own AOI (ranging from -0.05m2/ha for the LSMFI observatory to 0,55 m2/ha for Deodatie) but increased markedly when applied over other datasets. The largest bias was found on the Motherhouse model applied over the NFI dataset (+9.65 m2/ha), which represents more than 30% of the mean Ba for the whole AOI.

### 3.3. *Proportion of Extrapolated Pixels*

Table 4 shows the proportion of extrapolated pixels classified as "Far" from the calibration domain. The NFI and LSMFI Observatory models, showed the lowest number of extrapolated pixels over both the AOI and the forest datasets, with overall mean values of 24,2 % and 23,6% pixels as being "Far" from the calibration domains (highly extrapolated). From the local models, the Mouterhouse forest presented the lowest percentage of "Far" pixels (20,4%) in contrast to the Donon forest which had the highest proportion (37%).

*Table 4: Proportion of extrapolated pixels (%) classified as "Far" from the calibration domains, when models are applied to each forest dataset of the AOI*

|  |  | MODELS (% of extrapolated pixels) | | | | | | |
|---|---|---|---|---|---|---|---|---|
| **FOREST** | **Nº Pixels** | **Deodatie** | **Deux lacs** | **Donon** | **Gueberschwihr** | **Mouterhouse** | **LSMFI** | **NFI** |
| *Deodatie* | 369 648 | 36,3 | 62,2 | 34,7 | 84,6 | 56,6 | 27,7 | 24,5 |
| *Deux lacs* | 6 519 | 36, 8 | 36,7 | 36,0 | 95,2 | 68,9 | 23 | 25,5 |
| *Donon* | 126 480 | 45,8 | 62,3 | 37,0 | 85,3 | 53,5 | 29,1 | 26,8 |
| *Gueberschwihr* | 4 616 | 27,9 | 63,6 | 19,7 | 26,2 | 14,6 | 12,9 | 8,8 |
| *Mouterhouse* | 58 047 | 50,7 | 78,5 | 38,0 | 64 | 20,4 | 25, 2 | 14,1 |
| *AOI* | 3 917 489 | 43,6 | 72,7 | 41,2 | 74,2 | 41,5 | 29,1 | 21,1 |
| **Mean** |  | 40,90 | 67,8 | 33,9 | 80,7 | 47,0 | 23,6 | 24,2 |

Figure 5 illustrates the impact of sampling intensity on the proportion of extrapolated pixels using both the NFI and LSMFI observatory datasets. Over the AOI, it showed that the proportion of extrapolated pixels remained very high (>50%) even with a sampling intensity corresponding to 100-150 callibration plots (i.e. 1 plot every 1700 to 2500 ha).To complement the previous result, the relative RMSE and mean bias were computed against the sampling intensity for the NFI and LSMFI models combined over the AOI (Figure 5). A reduction in RMSE was acheived with such increase in sampling intensity. With an effort of ca. 200 to. 400 plots (i.e. 1 plot every 600 to 1200 ha), RMSE varied between 32-33%, and mean bias was neglectable (-10% to -20 %). A similar reduction occured with the mean extrapolation distance which remains high at low sampling intensities.

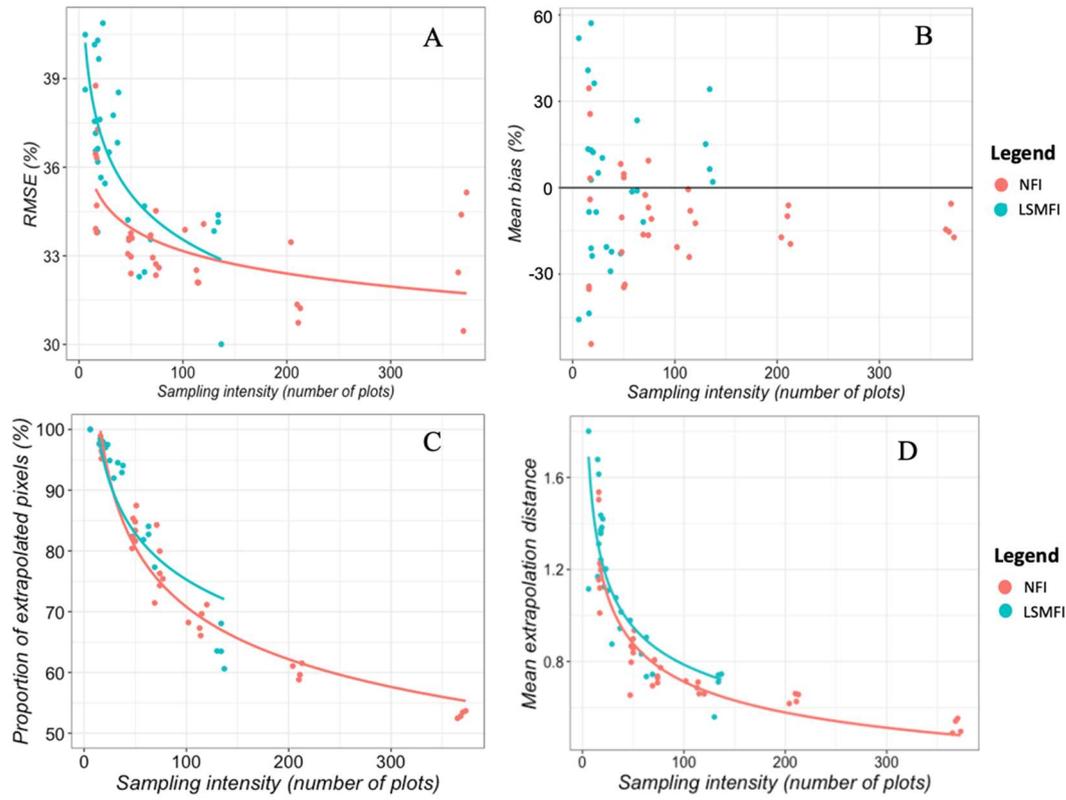

*Figure 5: Goodness of fit and extrapolation for LSMFI and NFI models compared to the sampling intensity. In A: the RMSE (%) ; B : Mean Bias (%) ; C : Proportion of extrapolated pixels ; D : Mean extrapolation distance*

Once the model's predictions and classifications of extrapolated pixels were computed, it was possible to generate maps of both predictions and associated risk ("Far" from the hull).

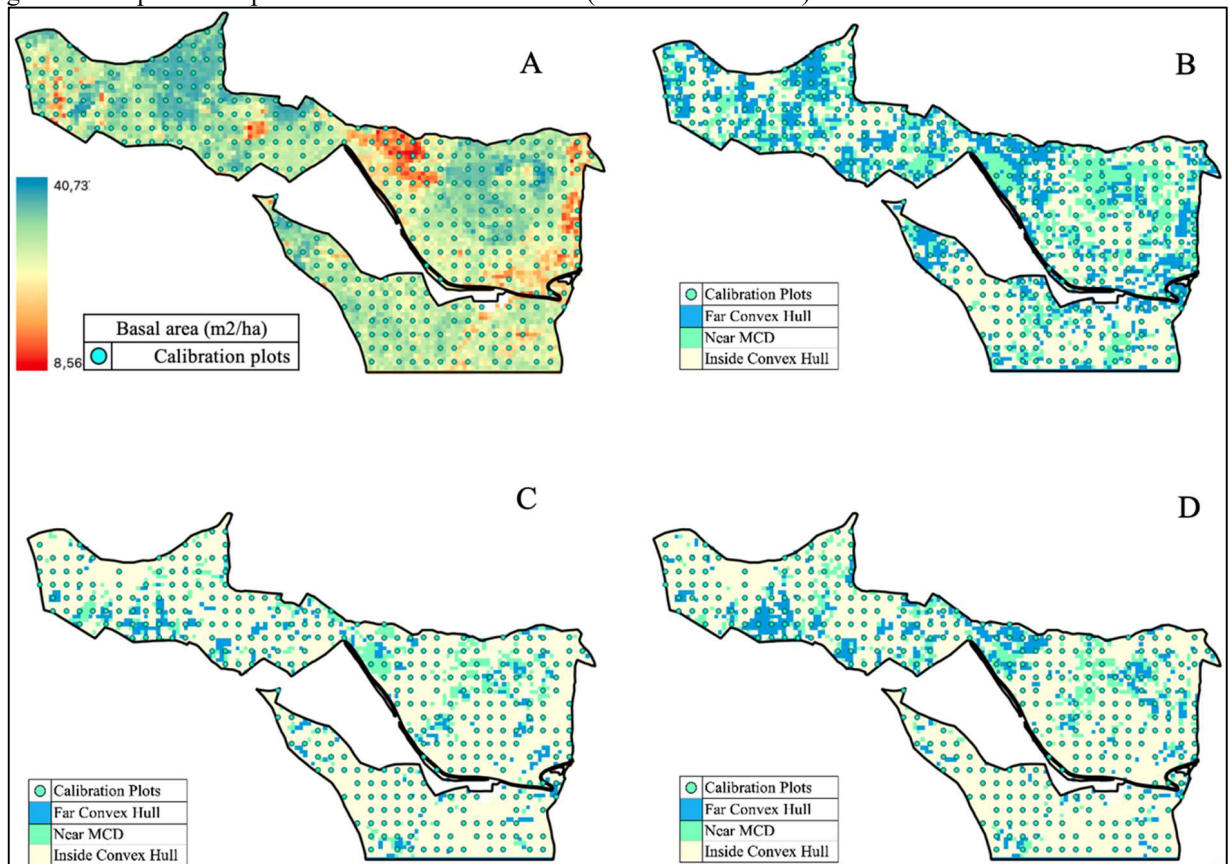

Figure 6 illustrated such results for the Gueberschwhir forest. Figure 6-A showed the spatial repartition of the predicted Ba: (m$^2$/ha) for the local model (the one developed using the plots illustrated). Ba in this forest ranged from 8,6 m$^2$/ha to 40,7 m$^2$/ha, with well identified spatial minima and maxima. Figure 6-B illustrated the classification of the extrapolated pixels still using the Gueberschwhir model. Several extrapolated pixels were "Far" from the hull, showing several spatial aggregates (dark blue). Figure 6-C and Figure 6-D illustrated the same results with models developed using the NFI and LSMFI observatory data respectively. Whereas many aggregates are still present, these later models exhibited a lower amount of largely extrapolated pixels (Far from the calibratin domain). The LSMFI observatory and the NFI models were presenting respectively 12.9% and 8.8% of pixels classified as Far. Among the local models, the Mouterhouse's one presented the lowest percentage of "Far" pixels (20,4%) in contrast to the Donon forest which had the highest proportion (37%) (Table 4).

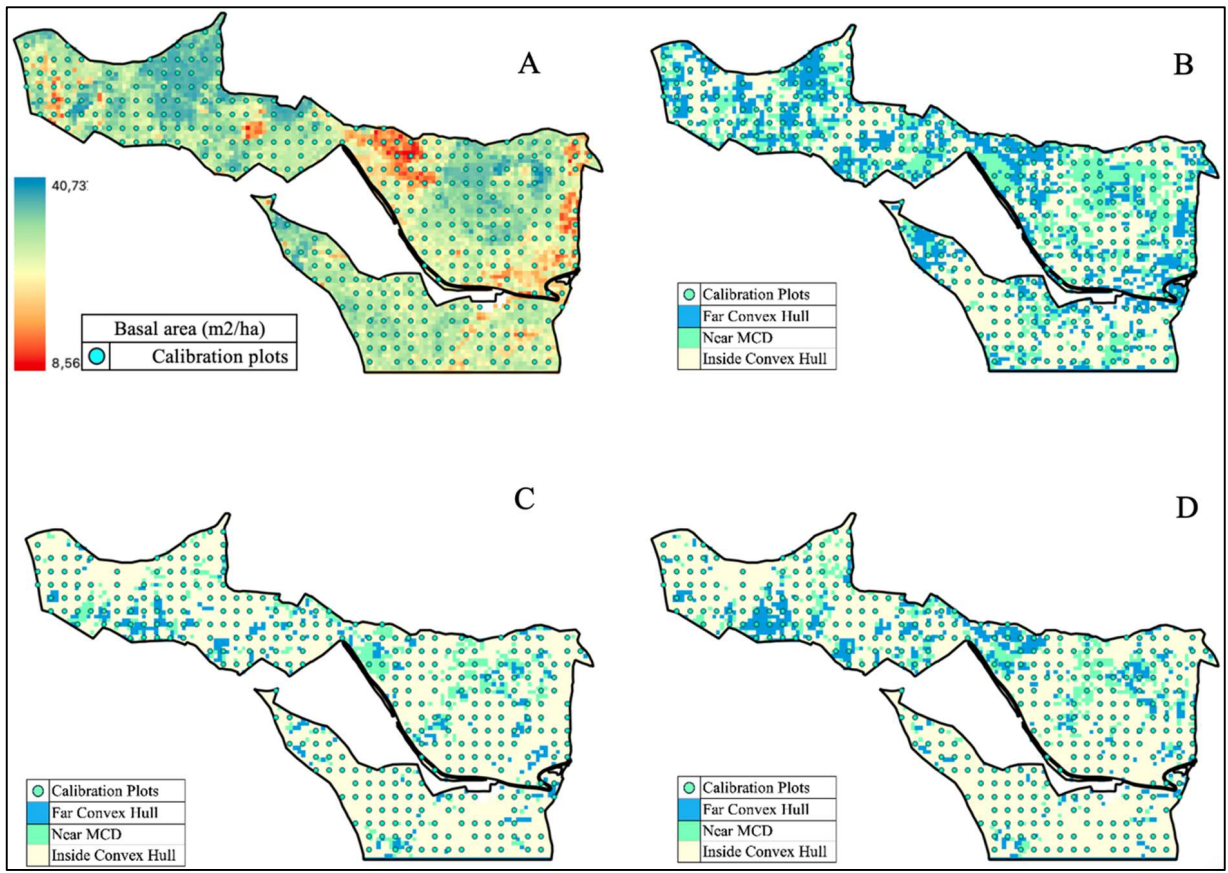

*Figure 6: Map products example for the forest of Gueberschwhir, In A= Basal area (m²/ha) prediction using the Gueberschwhir local model (The limits of forest and the network of calibration plots used in the model are shown); in B= pixels classification according to their extra polation classes for the same model. in C= pixels classification according to their extrapolation classes for the NFI model; and D=pixels classification according to their extrapolation classes for the LSMFI observatory model.*

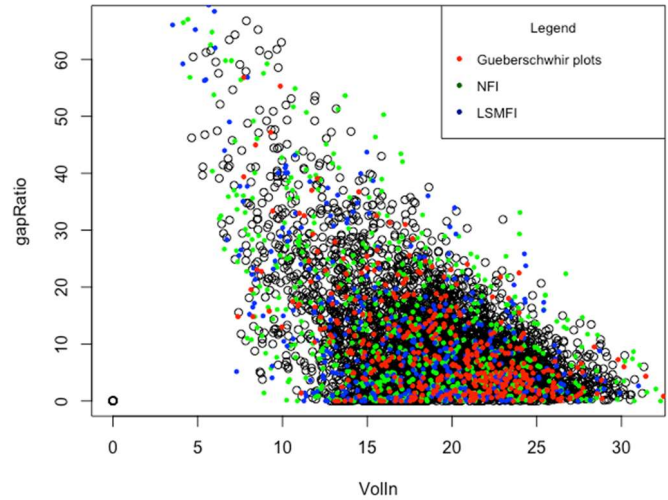

*Figure 7: Pixel coverage on the Gueberschwhir forest by the NFI, LSMFI observatory, and MFI plots on two LIDAR metrics : GapRatio and VolIn. In black : The pixels within the Gueberschwhir forest ; In green : NFI ; in blue :LSMFI and, red :MFI*

Figure 7 Illustrates the coverage of the inventory plot networks (NFI, LSMFI, MFI) on the Gueberschwhir forest for two metrics used in the model. While the NFI and the LSMFI networks cover a large proportion the spread of the whole pixel population shown in the figure, the MFI plots seem to have a more restricted distribution. This illustration gives an idea about how such big scale sampling networks could still contribute to model predictions over a specific AOI.

## 4. *Discussion*

We evaluated transferability and extrapolation risks for 7 RF models applied over independent datasets. The transferability indicators were $R^2$, RMSE, and Bias. Maps of Ba were produced as well as maps of extrapolated predictions classified as "Near" or "Far" according to their distance to their calibration domain. For the regional networks (NFI and LSMFI observatory) the impact of sampling efforts was also evaluated trough thining using iterations of the samples.

### *4.1. Model's transferability*

A notable aspect for each model was their good performance under local conditions. In general, local models explained over 60% of the Ba variability, with low RMSE and neglectable bias (Table 3). According to Guerra-Hernández et al. (2022), in model-based approaches the success of the model resides in the similarities between the sampled variable of interest and the domain to be estimated. However, when local models were applied to other datasets, the performance indicators rapidly deteriorated. According to Bartley et al. (2019), predictions done with limited data, and used to predict forest attributes at large scales is an issue. As seen on Table 1, the plot datasets differed among each other in terms of forest types. The models of Deux lacs and Deodatie for example, were build mainly with coniferous dominated plots. Deux Lacs had even no plots dominated by broadleaves, which may interfere in its transferability to broadleave or mixed forests (such as Donon, Gueberschwhir and Mouterhouse forests). As reported by Guerra-Hernández et al. (2022), when large samples are selected within the whole spectrum of forest types, more precise predictions are obtained. De Lera Garrido et al. (2023) also observed that diversity in forest conditions could affect, predictions quality locally and led to substaential errors and bias (>30% for basal area and volume).

Fekety et al. (2018), evaluated the transferability of RF models to predict Ba with ALS auxiliary data (mean height of the point cloud, percentage of ALS returns, and mean DTM elevation among others) and obtained relative RMSE ranging between 32% and 50%. This encompasses with the results obtained in this study where the relative RMSE for the regional plot networks from NFI and LSMFI observatory presented values around 32% to 33%. The authors suggested that for accurate predictions, it is important to apply models in management frameworks to forest with similar ecological features when working in unsampled areas.

In addition, Fekety et al. (2018) and Yates et al.(2018) reported that the number of plots, the attribute of interest, and the spatio-temporal variation of the data acquisition should receive a special attention during the development of models, as they influence directly goodness of fit. Regarding the regional networks of NFI and LSMFI, as they came from systematic sampling approaches over the whole AOI, they were covering a wider range of plot variability; this aspect was an advantage in terms of model transferability since its extents the validity domain of the models built from these datasets. However, the auxiliary data for the creation of the models came from several sources (ALS flight, Sentinel) which mismatched in acquisition time, and consequently could have influenced the quality of the results. Even though global networks covered a larger extent of the auxiliary space, predictions were found to be extrapolated for a substaential part of the AOI (Figure 5) and models could then be locally of low precision (Table 3). Ståhl et al. (2024) also underlined that magagers are often concerned by a specific portion of the population (e.g. old grown forest) that may not represent the average attribute measured in the calibration sample and could therefore be predictied with local biases.

According to Tompalski et al. (2019), the modelling approach and the level of similarity between forests, the target attribute and the auxiliary predictors are decisive in model's transferability. While evaluating transferability of models build from a linear regression, RF and a k-nearest neighbour approach, they observed that out of these 3 techniques, RF was found to be more stable for the prediction of forest attributes, with lower RMSE values. However, RF cannot extrapolate, which could be considered as a weakness of the method, especially when a low sampling intensity occured within an heterogenous AOI and when a substantial part of the predictions was out the calibration domain (e.g. as juged using convex hulls).

Table 3 also showed that for some models, $R^2$ appended to be negative when predictions were performed over other forests. The worst case occurred to be for the Deodatie's model used over the Mouterhouse forest ($R^2$ = -82%). This suggested that the predictions of the Deodatie model were worst that a direct simple mean of the Mouterhouse plots. These predictions were also largely biased (-9 m²/ha). This situation (very low or even negative $R^2$) indicated that models fited poorly to the data and therefore maps produced over these areas may be considered as potentially unreliable. This result was linked to extrapolation issues, since Table 4 also showed very high level of extrapolation in these cases.

An increase in sampling effort could diminish this problem, but it will always depend upon the level of similarity of the forests, for avoiding extrapolation. In this respect, the use of convex hull represents an interesting tool to evaluate the extent to which a calibration space is covering the auxiliary space of the poputation. An example where sampling intensity could be inefficient is for the Deux lacs forest, which has a high sampling intensity (1 plot every 2 ha), but any models build over this forest will always predict poorly broadleaves forest, since this forest type is absent from Deux Lacs. Another example is with the NFI model used over Mouterhouse. It has a $R^2$ of -27%. In this case, it is possible that the sampling intensity of the NFI network could be to low and may not cover the whole spectrum of of forest types observed in Mouterhouse, eventhough the NFI calibration domain seems to cover the whole AOI with only 14% of extrapolated pixels being "Far" from the calibration domain. In fact, in Mouterhouse, no NFI calibration plots were present, which could have exacerbated this transferability problem. This type of situation is illustrative of gaps within the calibration domain as explained by Meyer and Pebesma (2021).

### 4.2. Sampling effort

The sampling intensity plays an important role in model predictions. Our results showed that with a sampling effort of ca. 500 plots from regional networks (1 plot every 435 ha), there was a considerable reduction in the extrapolated predictions compared to a sampling effort of 127 plots (1 plot every 2000 ha) (Figure 5C). Extrapolation raised from ca. 50% to 73% respectively. Therefore, with a lower sampling effort and an increased variability in features of the forest charachteristics such as the species compostion, the proportion of extrapolated predictions rapidly increased, affecting the model's goodness of fit indicators ($R^2$, RMSE), and bias. In New Zealand,Tomppo et al. (1999) highlighted the need of establishing more field plots at local scales in small areas to provide more accurate estimates (between 20 to 50 or more plots in a block of 1000 ha of *Pinus radiata* plantations) to reach desired levels of precision. Additionally, De Lera Garrido et al. (2023) in an assessment of nationwide forest attribute maps of Norway, mentioned how the sample size could led to systematic errors when it comes from NFI or LSMFI networks. Large networks are not designed to account for all particularities of smaller AOIs.Then, their accuracy is reduced compared to local FMI's. Consequently, with a large sample size to calibrate models, it is possible to expand the validity domain, and improve model transferability.

An important highlight to consider, could also be the implementation of Local Pivotal Sampling (LPS) methods, to reduce the levels of extrapolation when a generic model is used with a known auxiliary space (Grafström et al. 2014). According to Brus (2022), the advantage of LPS consists in making proportional sampling units according to the area occupied by the target attribute within the sampling grid cells, thus it could improve the coverage of the auxiliary space by the calibration domain. Unfortunately, the models are rarey known before hand and the field work is often started prior to the remote sensing data acquisition, making the LPS approach less attractive.

### 4.3. The validity domain and extrapolation risks

As stated by several authors (Brooks et al. 1988; Renaud et al. 2022), the convex hull approach is a convenient and practical way to easily establish the validity domain of a model in terms of its auxiliary space as far as the models used are parcimonious (i.e. less then 9 variables). The regional networks (NFI, LSMFI) showed the lowest amount of potentially unreliable pixels classified as "Far" from the calibration domain. On the maps, these pixels were often spatially aggregated and should be considered with special attention since they could represent unreliable predictions (Bartley et al. 2019; Renaud et al. 2022; Sagar et al. 2022).

In our study, when models were applied locally, the percentage of extrapolated pixels remained low in comparison to local models transferred to other forest datasets as seen in table 4. From the total pixels over the AOI (3 917 478 pixels) with NFI and LSMFI models, the mean proportion of extrapolated pixels classified as "Far" from their calibration domain were 24,2% and 23,6% for the NFI and LSMFI models respectively. According to Zurell et al. (2012) understanding the extrapolation behaviour is a concern in model's transferability to unobserved areas. Therefore producing maps of extrapolation levels, may be a good practice for forest management (Tomppo et al. 1999; Sagar et al. 2022), as it could avoid using potentially erroneous predictions. Coops et al.(2021) emphasises that extrapolation risks are prone to happen over large areas when there is not enough representativity of the plots describing the complete range of the forest attribute of interest. Fekety et al.(2015) adds that as the density of plots increases, the chances to improve model calibration also increases. Unfortunately, no sampling rules are currently available prescribing a sampling scheme for any AOI, based on a forest heterogeneity index for example. Such rules though could help improving quality of remote sensing maps performed over large AOI.

*4.4. Extrapolation maps for forest management*

Another important aspect of this study was to produce "reliability" maps based on pixels' classification of extrapolation levels. As seen in Figure 6, for the Gueberschwhir forest, the spatial representation of the potentially unreliable predictions may warn forest managers about doubtful model predictions that merit verifications. With the convex hull method, it is easy to spot these areas. According to Bartley et al. (2019), with an insufficient sample, a biased representation of the population could often occurs. Additionally Meyer and Pebesma (2021) as well as Kangas et al. (2023) emphazised the importance of spotting uncertain spatial predictions in orther to increase the confidence on large-scale maps. Figure 5 is therefore an interesting tool providing forest owners a way to improve their decision-making processes.

Frank et al. (2020) indicated how forest inventories combined with remote sensing data can support management decisions, while paying attention to uncertainties and the scale under which these decisions are to be taken. NFI and LSMFI observatory models, as they cover a large range in forest variability are reducing extrapolation risks associated to model predictions. However, special attention should be kept, as the performance indicators of the NFI and LSMFI observatory models still present locally major warnings in terms of RMSE/Bias and less predictive power for Ba, as can be seen for example in Deodatie or Deux lacs FMIs (Table 3). One way to improve this type of extrapolation assessment according to Fekety et al. (2018) is to limit predictions to more homogeneous type of stands, in order to use comparable auxiliary information in model building and predictions. De Lera Garrido et al. (2023) however have not observed any model improvements when a stratification step was applied to their NFI data in Norway. A way to further improve global model locally, when issues are important, would be to complement locally the sampling design, as suggested by Tomppo et al. (1999). Meanwhile, Meyer and Pebesma (2021) also highlighted the issue behind the use of machine learning approaches such as RF, wich are data-driven and are laking the possibility to extrapolate. Another approach, proposed by Kangas et al (2023), to provide uncertainty assessment at the pixel level would be to use Kiging methods. The reliability associated to model predictions and maps uncertainties remain an important aspect to be further studied.

## 5. Conclusion

An approach was presented to evaluate models' transferability and risks of potentially unreliable predictions. Results showed that local models build over a dense local network of management plots are performing well at their local scales, but are at risk when transferred to other AOIs. Conversely, global models build upon regional networks may cover a larger range of the auxiliary space in term of calibration domain, but may produce biased predictions locally when applied to forests that diverge from their model's validity domain. These local reliability problems appeared to be frequent, as it was observed in every forest tested in this study. The regional network examined appeared to be sensitive to extrapolation problems, since more than 50% of the predictions were outside the model's calibration domain in the AOI examined. This result thus emphasised the need to consider local bias when models are build from regional networks with a relatively low sampling intensity.

The extrapolated predictions could easily be identified and mapped over an AOI usging a convex hull approach. Highlighting potentially unreliable predictions, may help forest managers to take informed decisions with the maps presented. However for high-dimension models, convex hull is not a practical solution and other approaches would have to be considered to evaluate the extrapolation level of models with the aim of producing locally unbiased maps.